\documentclass{ws-procs975x65}

\def\figsubcap#1{\par\noindent\centering\footnotesize(#1)}

\begin{document}

\title{\uppercase{Less is more: How cosmic voids can shed light on dark energy}}

\author{\uppercase{E. G. Patrick Bos}$^*$, \uppercase{Rien van de Weygaert} and \uppercase{Jarno Ruwen}}

\address{Kapteyn Astronomical Institute, University of Groningen,\\
Groningen, 9747 AD, The Netherlands\\
$^*$E-mail: pbos@astro.rug.nl\\
www.rug.nl/sterrenkunde}

\author{\uppercase{Klaus Dolag}}
\address{Department of Physics, Ludwig-Maximilians-Universit\"at, Scheinerstr. 1,\\
M\"unchen, 81679, Germany}

\author{\uppercase{Valeria Pettorino}}
\address{D\'epartement de physique th\'eorique, Universit\'e de Gen\`eve, 24 Quai Ansermet,\\
Gen\`eve, 1211, Switzerland}

\begin{abstract}
We showed how the shape of cosmic voids can be used to distinguish between different models of dark energy using galaxy positions.
\end{abstract}


\bodymatter

\section{Goals}
\begin{figure}[t]
\begin{center}
\psfig{file=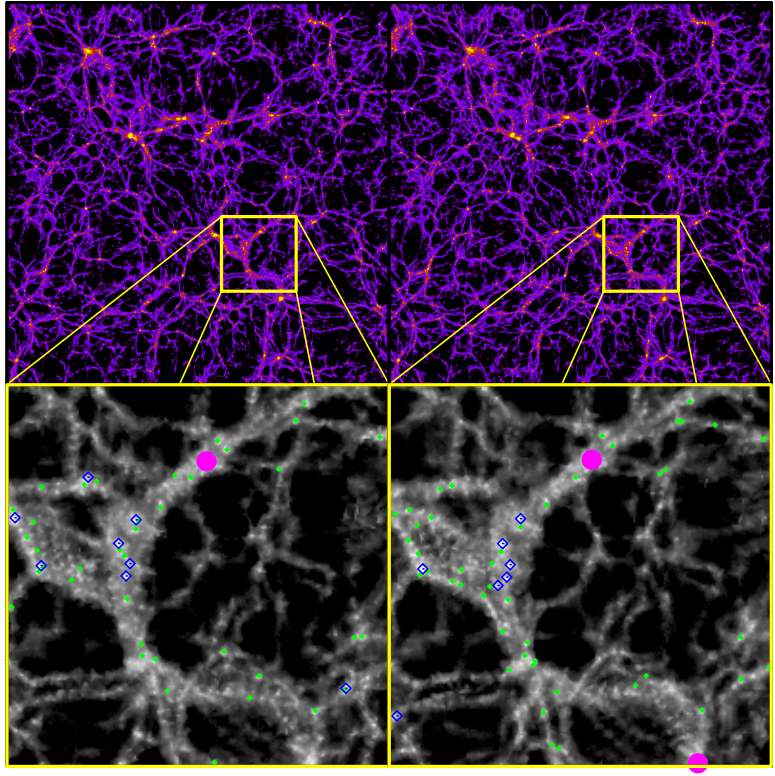, width=0.8\textwidth}
\end{center}
\caption{$\Lambda$CDM (left) and quintessence (right) cosmologies; dark matter density (grey) and galaxies (coloured dots). Differences are subtle; measuring them is the challenge presented here.}
\label{fig0}
\end{figure}
In Ref.~\refcite{bos12} we assessed the sensitivity of void shapes to the nature of dark energy that was pointed out
in recent studies \cite{parklee07, leepark09, lavaux10}. We investigated whether or not void shapes are useable
as an observational probe in galaxy redshift surveys. We focused on the evolution of the mean void ellipticity and
its underlying physical cause. 

\section{Methods}
\begin{figure}[b]
\begin{center}
  \parbox{0.48\textwidth}{\epsfig{figure=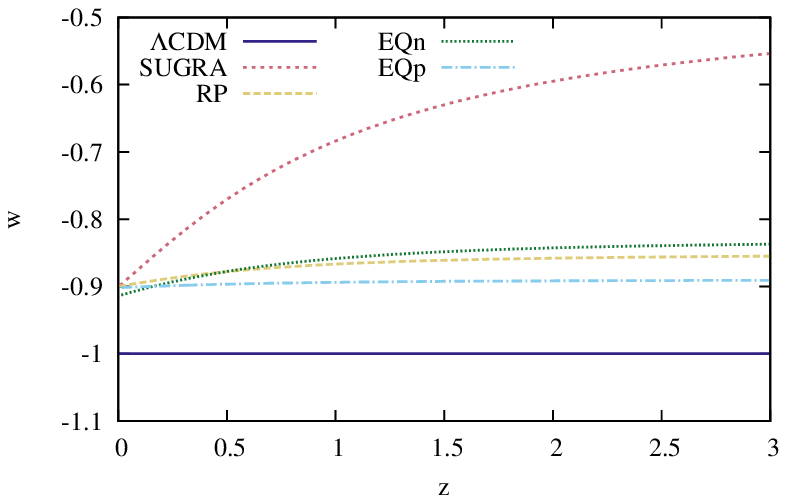,width=0.48\textwidth}\figsubcap{a}}
  \hspace*{0.01\textwidth}
  \parbox{0.48\textwidth}{\epsfig{figure=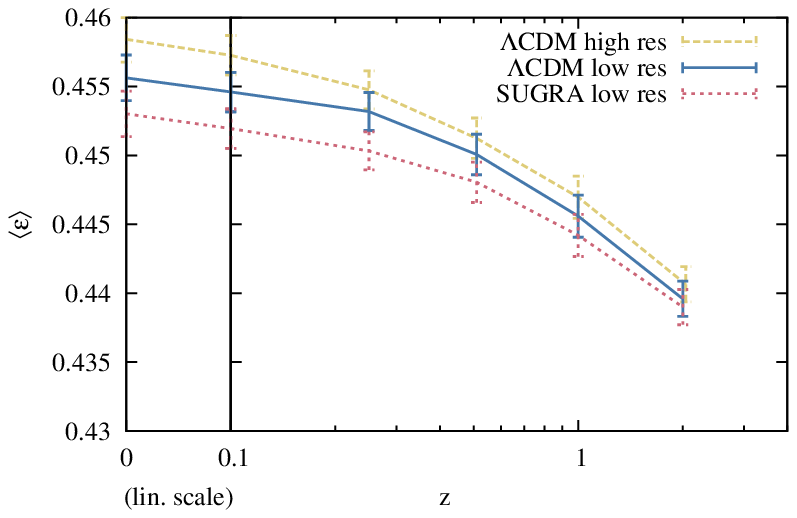,width=0.48\textwidth}\figsubcap{b}}
  \caption{(a) Equation of state parameter $w$ versus redshift. Shows the evolution of the dark energy equation of state parameter $w = P/\rho$. The different lines indicate different dark energy models. (b) Mean ellipticity as a function of redshift, determined from the full dark matter particle distribution using the WVF/DTFE scheme. $\Lambda$CDM high and low resolution simulations and a SUGRA low resolution one.}
  \label{fig12}
\end{center}
\end{figure}

To this end, we analysed the morphological properties of voids in five sets of cosmological N-body
simulations\cite{deboni11}, each with a different nature of dark energy, as expressed by the equation of state parameter $w$,
that relates the pressure of the dark energy cosmic fluid to its density through $P = w \rho$ (figure \ref{fig12}.a).
Comparing voids in the dark matter distribution to those in the halo population, we addressed
the question of whether galaxy redshift surveys yield sufficiently accurate void morphologies.

At first, voids are identified using the parameter free Watershed Void Finder (WVF)\cite{platen07}, applied on a density field
obtained by means of a Delaunay Tesselation Field Estimator (DTFE)\cite{schaap00, cautun11a}.
Secondly, we tried using a different void finder, based on the Monge-Amp\`ere-Kantorovitch (MAK) reconstruction
of Ref.~\refcite{lavaux10}. This method yields a different type of void that may be sensitive to the underlying cosmology
in a different way than the WVF/DTFE method.

The effect of redshift distortions was investigated as well. The distorting effect of these on the shape measurements are
included in the figures as error-bars.

\section{Results}
We confirmed the statistically significant sensitivity of voids in the dark matter distribution (figure \ref{fig12}.b).

\begin{figure}[b]
\begin{center}
  \parbox{0.48\textwidth}{\epsfig{figure=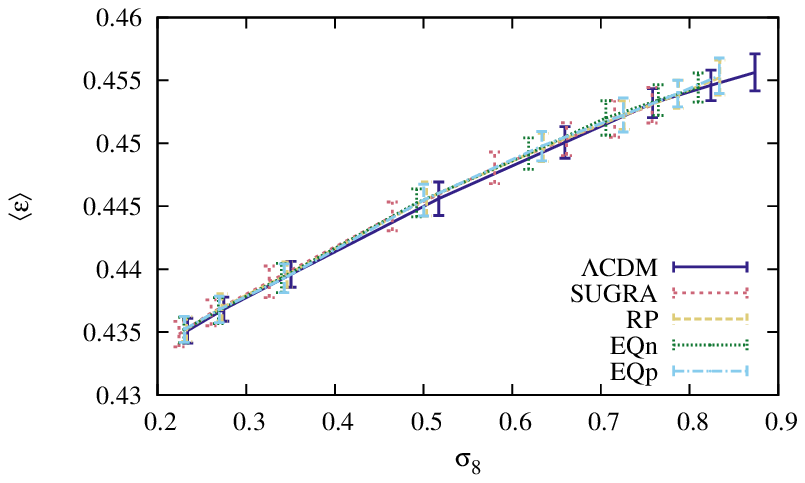,width=0.48\textwidth}\figsubcap{a}}
  \hspace*{0.01\textwidth}
  \parbox{0.48\textwidth}{\epsfig{figure=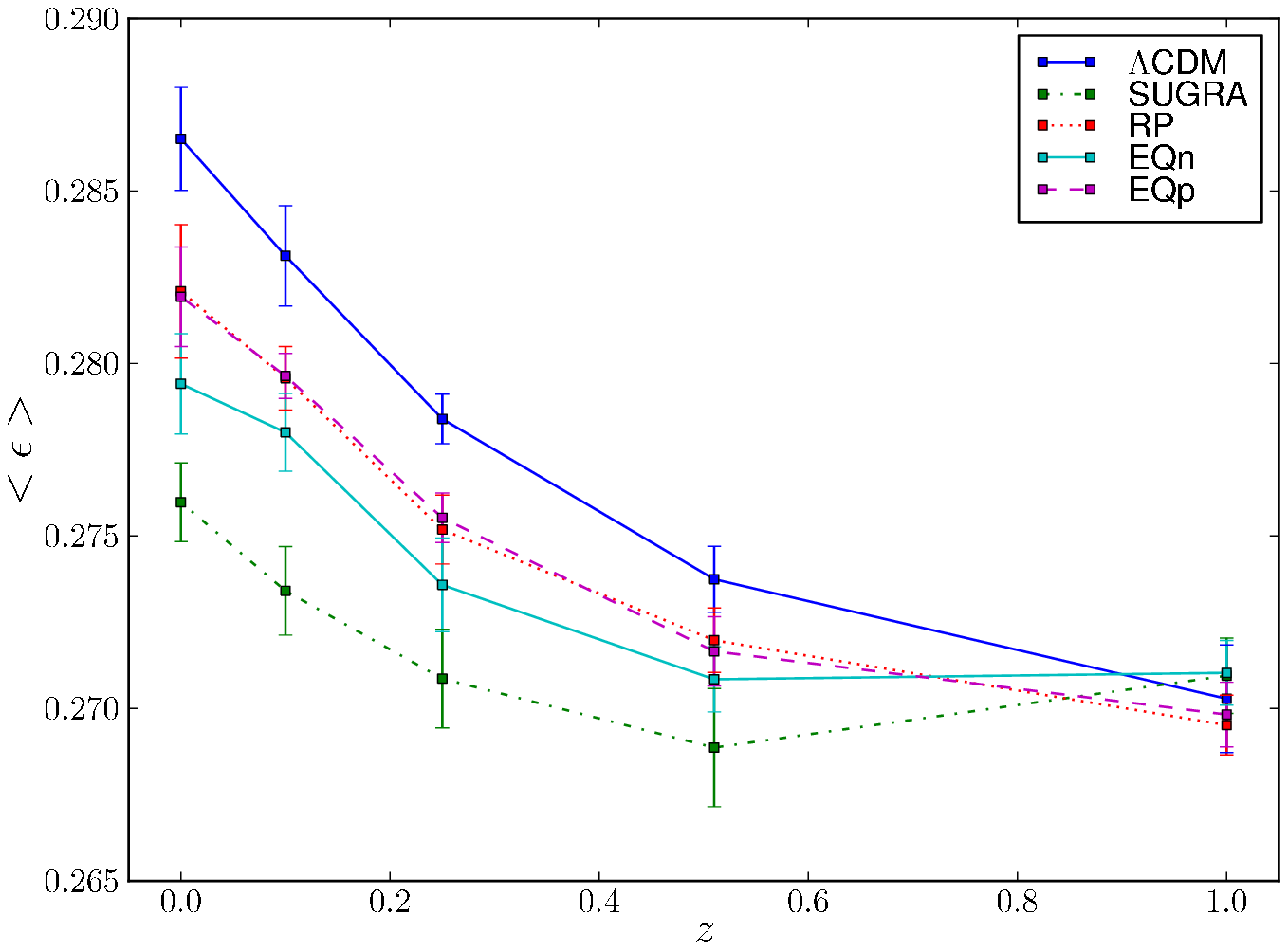,width=0.48\textwidth}\figsubcap{b}}
  \caption{(a) Mean ellipticities versus $\sigma_8$. The different lines show the low resolution simulations of the five different cosmologies. The lines consist of $\sigma_8$ values at redshifts $0$, $0.1$, $0.25$, $0.51$, $1.0$ and $2.04$. (b) Again, mean ellipticity as a function of redshift, but now determined from the halo/galaxy distribution using the MAK void finder. Clearly, the differences between models are significantly discernible using this setup.}
  \label{fig34}
\end{center}
\end{figure}

The level of clustering as measured by $\sigma_8(z)$ was identified as the main cause of differences in mean void
shape $\langle\epsilon\rangle$ (figure \ref{fig34}.a).

We further found that using the combination of WVF with DTFE in the halo and/or galaxy distribution it is practically
unfeasible to distinguish at a statistically significant level between the various cosmologies due to the sparsity
and spatial bias of the sample.

However, using the MAK void finder, the (mock) galaxy distribution yields excellent results; the effects of
different dark energy models are clearly discernible in the evolution of the mean ellipticity (figure \ref{fig34}.b).
This is a promising result in the light of upcoming galaxy redshift surveys, on which this method may be applied.

\bibliographystyle{ws-procs975x65}

\end{document}